\documentclass[epj]{svjour}
\usepackage{graphics}
\usepackage{amsmath}
\usepackage{xcolor}
\usepackage{braket}
\usepackage{cases}
\usepackage{slashed}
\usepackage[utf8]{inputenc}
\usepackage[T1]{fontenc}
\usepackage{ulem}

\begin{document}
\title{Nuclear response theory for spin-isospin excitations in a relativistic quasiparticle-phonon coupling framework}
\author{Caroline Robin \and Elena Litvinova
}                     
\institute{Department of Physics, Western Michigan University, Kalamazoo, MI 49008-5252}
\date{Received: date / Revised version: date}
%
\abstract{
A new theoretical approach to spin-isospin excitations in open-shell nuclei is presented. The developed method is based on the relativistic meson-exchange nuclear Lagrangian of Quantum Hadrodynamics and extends the response theory for superfluid nuclear systems beyond relativistic quasiparticle random phase approximation in the proton-neutron channel (pn-RQRPA). The coupling between quasiparticle degrees of freedom and collective vibrations (phonons) introduces a time-dependent effective interaction, in addition to the exchange of pion and $\rho$-meson taken into account without retardation. The time-dependent contributions are treated in the resonant time-blocking approximation, in analogy to previously developed relativistic quasiparticle time blocking approximation (RQTBA) in the neutral (non-isospin-flip) channel. The new method is called proton-neutron RQTBA (pn-RQTBA) and applied to Gamow-Teller resonance in a chain of neutron-rich Nickel isotopes $^{68-78}$Ni. A strong fragmentation of the resonance along with quenching of the strength, as compared to pn-RQRPA, is obtained. Based on the calculated strength distribution, beta-decay half-lives of the considered isotopes are computed and compared to pn-RQRPA half-lives and to experimental data. It is shown that a considerable improvement of the half-life description is obtained in pn-RQTBA because of the spreading effects, which bring the lifetimes to a very good quantitative agreement with data.
} 
\authorrunning{C. Robin \and E. Litvinova}
\titlerunning{Spin-isospin excitations in a relativistic quasiparticle-phonon coupling framework}
\maketitle
\section{Introduction}
\label{intro}
Charge-exchange excitations of nuclei are crucial in our understanding of many phenomena extending across the fields of nuclear physics and astrophysics. The study of such modes of excitations can, for instance, constrain the spin-isospin part of the nuclear interaction. 
Well-defined model-independent sum rules allow for indirect determination of neutron and proton radii \cite{HW.GR,Krasznahorkay} and provide information on the equation of state \cite{Typel}. In addition, these types of modes determine the rates of weak interaction processes, such as beta-decay, neutrino scattering or electron capture, occurring in astrophysical environments. In particular, the low-lying part of the Gamow-Teller transition strength provides the beta-decay half-lives which, together with the neutron-capture rates, govern the r-process nucleosynthesis. Over the last years, this field of research has become extremely active, as new experimental facilities, such as FRIB \cite{FRIB} or FAIR \cite{FAIR}, will be able to provide in the time to come new data on exotic nuclei. Nevertheless, many of the nuclei of interest for the r-process remain too far away from the valley of stability to be reached experimentally in the near future. Thus, it is essential to be able to describe in a microscopic way isospin-transfer nuclear excitations, in order to provide an accurate and consistent input for the modeling of stellar evolution. 
Over the past, an important number of theoretical works dedicated to the study of these types of excitations have been realized within various approaches. They range from ab-initio methods \cite{Navratil} to Shell-Model techniques \cite{Brown,Caurier,Koonin}, and methods based on the extensions of the self-consistent mean-field, each with different domains of applicability. Within the latter type of approaches, many investigations have been done using the so-called Random-Phase Approximation (RPA), and its extension to superfluid systems, the Quasiparticle RPA (QRPA). Starting from a non-relativistic G-matrix \cite{Fang}, Skyrme \cite{Engel,Sarriguren}, Gogny \cite{Martini} or Fayans \cite{Borzov} mean-field, or from a relativistic Hartree \cite{Niksic} or Hartree-Fock (HF) field \cite{Liang}, the (Q)RPA is able to calculate transition probabilities up to high excitation energies, and has usually been quite successful in reproducing the position of the giant resonances. However, it is a well-known problem that this approach is unable to describe the spreading width of the resonances. As it only accounts for one particle-one hole (1p1h) or two quasiparticle (2qp) excitations on top of the ground state, the (Q)RPA transition strength is usually found to be concentrated in a single or a few peaks (mechanism known as Landau damping).
In particular, the experimental Gamow-Teller (GT) strength, when measured up to around the giant resonance energy, generally only reproduces about $70\%$ of model-independent Ikeda sum rule, while the (Q)RPA usually exhausts almost the full sum rule within the same energy window.
Besides the possibility of exciting a nucleon into a $\Delta$ resonance, the main contribution to this "quenching" of the GT distribution is thought to be due to higher-order configurations, more complex than 1p-1h or 2qp, which are expected to fragment and shift a part of the strength to higher energy. In the absence of such effects, (Q)RPA also usually systematically overestimates the resulting beta-decay half-lives. A solution to this discrepancy is often found by introducing an isoscalar proton-neutron pairing, whose parameter is adjusted separately for each isotopic chain \cite{Niksic}. Approaches beyond the (Q)RPA include 2p-2h or 4qp configurations, in order to reach a better description of Gamow-Teller transitions and associated lifetimes in a more microscopic and satisfying way. Second RPA for isospin-flip excitations was realized in Ref. \cite{Drozdz}, where a comprehensive investigation of spectra of various multipolarities in doubly-magic nuclei has quantified fragmentation effects beyond RPA. More recently, Ref. \cite{Severyukhin} studied the influence of phonon-phonon coupling on the description of $\beta$-decay rates, based on a Skyrme HF+BCS mean-field and a residual Landau-Migdal force. Finally, a lot of progress has been made in the past few years towards the development of self-consistent approaches based on the particle-vibration coupling (PVC) model. In this framework the coupling between the motion of single-nucleons and the collective vibrations (phonons) is considered, so that 2p2h configurations
in the 1p1h$\otimes$phonon coupling scheme, are explicitly taken into account. A non-relativistic version of the PVC approach has led to an overall good description of GT resonances and half-lives, using various Skyrme forces \cite{Niu2012,Niu2014,Niu2015}. The so-called Relativistic Time-Blocking Approximation (RTBA) has also quite successfully investigated the effect of the particle-vibration coupling on Gamow-Teller and spin-dipole modes in a relativistic framework \cite{Marketin,Litvinova2014}. As no pairing correlations were implemented, this approach was, however, limited to magic nuclei. 
\\
In this work we extend the RTBA method to the description of spin-isospin excitations in open-shell nuclei. The resulting approach is a combination of the Relativistic Quasiparticle Time Blocking Approximation (RQTBA) that was developed in the neutral channel \cite{Litvinova2008} and the proton-neutron RTBA (pn-RTBA). In what follows it will be referred to as pn-RQTBA. 
In section \ref{sec:formalism}, we present in detail the formalism of the pn-RQTBA. In section \ref{sec:appli}, we apply this method to calculations of Gamow-Teller transitions in neutron-rich Nickel 
isotopes. This isotopic chain is of particular interest for astrophysics as it plays an important role in the r-process nucleosynthesis. The strength distributions are then used to calculate $\beta$-decay half-lives which are compared to experimental data. Finally, we give conclusions and perspectives to this work in section \ref{sec:conclu}.

\section{Formalism}
\label{sec:formalism}
In this section we present the formalism of the pn-RQTBA approach, which is obtained by combining the neutral-channel RQTBA presented in \cite{Litvinova2008} and the pn-RTBA introduced and applied in \cite{Marketin,Litvinova2014}. 
\subsection{One-nucleon motion in a relativistic framework}
In a standard relativistic theory of nuclei \cite{Ring}, the nucleus is governed by an effective Lagrangian describing the interaction between Dirac nucleons in terms of boson (meson and photon) exchange:
\begin{eqnarray}
\mathcal{L} &=& \mathcal{L}_{nucl} + \mathcal{L}_{bos} +\mathcal{L}_{int} \; ,
\label{eq:Lagrangian}
\end{eqnarray}
where 
\begin{equation}
\mathcal{L}_{nucl} =  \bar \psi (i \slashed \partial - m ) \psi \; ,
\end{equation}
is the free Lagrangian for the nucleon field $\psi$, $\bar \psi = \psi^\dagger \gamma^0$ and $m$ denotes the bare nucleon mass. $\mathcal{L}_{bos}$ is the free Lagrangian for the minimal set of effective mesons $(\sigma, \omega^\mu, \overrightarrow \rho^\mu)$ and electromagnetic field $A^\mu$:
\begin{eqnarray}
\mathcal{L}_{bos} &=&  \frac{1}{2} \partial_\mu \sigma \partial^\mu \sigma - U(\sigma) \nonumber \\
                   &&    - \sum_{b = \omega^\mu, \overrightarrow \rho^\mu, A^\mu} \left( \frac{1}{4} \mathcal{F}_{\mu\nu}(\phi_b) \mathcal{F}^{\mu\nu}(\phi_b) - U(\phi_b) \right)
\label{eq:Lag_bos}
\end{eqnarray}
where the field tensors read
\begin{numcases}{}
 \mathcal{F}_{\mu\nu}(\omega^\mu) \equiv \Omega_{\mu\nu} = \partial_\mu \omega_\nu - \partial_\nu \omega_\mu \nonumber \\
 \mathcal{F}_{\mu\nu}(\overrightarrow \rho^\mu) \equiv  \overrightarrow R_{\mu\nu} = \partial_\mu \overrightarrow \rho_\nu - \partial_\nu \overrightarrow \rho_\mu \nonumber \\
 \mathcal{F}_{\mu\nu}(A^\mu) \equiv  F_{\mu\nu} = \partial_\mu A_\nu - \partial_\nu A_\mu \;, \nonumber
\end{numcases}
where the arrow denote vectors in isospin space. In this work a density dependence, necessary for an accurate description of nuclear properties, is simulated by using a non-linear $\sigma$ self-interaction. 
The potentials in (\ref{eq:Lag_bos}) thus read
\begin{numcases}{}
 U(\sigma) = \frac{1}{2} m_\sigma^2 \sigma^2 + \frac{g_2}{3} \sigma^3 + \frac{g_3}{4} \sigma^4 \; ,\\
 U(\phi_b) = \frac{1}{2} m_b^2 \phi_b^2 \;, \; b=(\omega^\mu,\overrightarrow \rho^\mu, A^\mu) \; .
\end{numcases}
Finally, the interaction Lagrangian is given by
\begin{eqnarray}
 \mathcal{L}_{int} = - \sum_{b = \sigma,\omega^\mu,\overrightarrow \rho^\mu, A^\mu }  \bar \psi \Gamma_b \phi_b \psi \; ,
\end{eqnarray}
where the coupling vertices read
\begin{eqnarray}
\begin{matrix}
 \Gamma_\sigma = g_\sigma \;, & &\Gamma_\omega^\mu = g_\omega \gamma^\mu \; ,\\
 \overrightarrow \Gamma_\rho^\mu = g_\rho \overrightarrow\tau \gamma^\mu \;,&& \Gamma_A^\mu = e \frac{1 - \tau_3}{2} \gamma^\mu \; .
 \end{matrix}
\end{eqnarray}
\\
\\
The dynamics of a nucleon is in theory encapsulated in the one-body Green's function 
\begin{equation}
g(x_1,x_2) = -i \braket{0 |\mathcal{T} \left( \hat\psi(x_1) \hat{\bar\psi}(x_2)\right)| 0} \; ,
\label{eq:sp_GF1} 
\end{equation}
where $x_1\equiv (t_1, \mathbf{r}_1)$ is the space-time coordinate, and $\mathcal{T}$ is the time-ordering operator. The state $\ket{0}$ denotes in principle the "exact" ground state of the interacting system, and $\hat\psi(x_1)$ and $\hat{\bar\psi}(x_1)$ are the quantized destruction and creation nucleon field operators in the Heisenberg representation.  
\\
In approaches based on Covariant Density Functional Theory (CDFT) \cite{Ring}, one starts from the mean-field description of the system in order to determine a first approximation for the ground state of the nucleus, thus leading to first approximation for the one-nucleon propagator (\ref{eq:sp_GF1}). Complex configurations arising from correlations between the nucleons are then included in a second step.
\\
\\
In superfluid Fermi system, such as open-shell nuclei, it is necessary to include pairing correlations consistently at the unperturbed mean-field level and beyond. In the Green's function formalism this is commonly done by introducing, in addition to the normal propagator (\ref{eq:sp_GF1}), the so-called anomalous Green's functions allowing for the destruction and creation of pairs of nucleons. This can be expressed in a compact way, in the Nambu-Gorkov representation \cite{Gorkov,Nambu} which defines generalized operators of creation and annihilation of quasinucleon fields:
\begin{equation}
\hat{\boldsymbol{\Psi}}(x_1)=
\begin{bmatrix}
\hat\psi(x_1) \\
\hat{\bar\psi}^T(x_1)
\end{bmatrix}
\;, 
\hspace{0.2cm}
\hat{\bar{\boldsymbol{\Psi}}}(x_1)=
\begin{bmatrix}
\hat{\bar\psi}(x_1) & \hat\psi^T(x_1)
\end{bmatrix}
\; .
\label{eq:general_field}
\end{equation}
The corresponding Green's function becomes a 8x8 matrix that reads
\begin{eqnarray}
&&\mathbf{G}(x_1,x_2) = -i \braket{0 |\mathcal{T} \left( \hat{\boldsymbol{\Psi}}(x_1) \hat{\bar{\boldsymbol{\Psi}}}(x_2)\right)| 0} \nonumber \\
 &&               = -i 
\begin{pmatrix}
 \braket{0 |\mathcal{T} \left( \hat\psi(x_1) \hat{\bar\psi}(x_2)\right)| 0} & \braket{0 |\mathcal{T} \left( \hat\psi(x_1) \hat\psi^T(x_2)\right)| 0} \\
 \braket{0 |\mathcal{T} \left( \hat{\bar\psi}^T(x_1) \hat{\bar\psi}(x_2)\right)| 0}  & \braket{0 |\mathcal{T} \left( \hat{\bar\psi}^T(x_1) \hat\psi^T(x_2)\right)| 0} 
\end{pmatrix} 
\nonumber \\
&&                \equiv 
\begin{pmatrix}
G(x_1,+ ; x_2,+) & G(x_1,+ ; x_2,-)  \\
G(x_1,- ; x_2,+) & G(x_1,- ; x_2,-)
\end{pmatrix} 
 \; ,
\nonumber \\
\label{eq:GF1} 
\end{eqnarray}
where we have introduced an extra index $\chi = \pm$ to denote the different components in the doubled coordinate space $(x,\chi)$, as in Ref. \cite{Tselyaev}.
Note that the Heisenberg operators $\hat{\psi}(x_1)$ and $\hat{\bar\psi}(x_1)$ are now defined relatively to $\hat H' = \hat H - \lambda_N \hat N - \lambda_P \hat Z$, $\lambda_N$ and $\lambda_P$ being the neutron and proton chemical potentials respectively, and $\hat N$ and $\hat Z$ the neutron and proton number operator respectively:
\begin{numcases}{}
\hat{\psi}(x_1) = e^{iH't_1}  \hat{\psi}(\mathbf{r}_1) e^{-iH't_1}  \; , \\
\hat{\bar\psi}(x_1) =e^{iH't_1} \hat{\bar\psi}(\mathbf{r}_1) e^{-iH't_1} \; .
\end{numcases}
Also, $\ket{0}$ only has a good particle number on average, as it now denotes in principle the ground state of the interacting system governed by $\hat H'$. 
\\
\\
The propagator (\ref{eq:GF1}) is the solution of the Dyson equation
\begin{eqnarray}
\mathbf{G}(x_1,x_2) &=& \mathbf{G}^0(x_1,x_2) \nonumber  \\ 
                                && + \int dx_3 dx_4 \; \mathbf{G}^0(x_1,x_3) \mathbf{\Sigma}(x_3,x_4) \mathbf{G}(x_4,x_2) \; , \nonumber \\
\label{eq:Dyson1}
\end{eqnarray}
where $\mathbf{G}^0 (x_1,x_2) = \begin{pmatrix} g^0(x_1,x_2) & 0 \\ 0 & -g^{0T}(x_2,x_1) \end{pmatrix}$ is the free propagator solution of
\begin{equation}
 \left( i \slashed \partial_1 - m \right) g^0(x_1,x_2) = \delta(x_1-x_2) \times I_{4\times 4} \; ,
\end{equation}
and $\mathbf{\Sigma}$ is the so-called self-energy (or mass operator), which in principle resums all the possible processes that a quasinucleon can undergo when it propagates in the nuclear medium.
\\
\\
To determine the propagator $\mathbf{G}$, we split the self-energy into a static part, which includes only instantaneous processes, and an energy-dependent part, which describes the dynamics:
\begin{eqnarray}
\mathbf{\Sigma}(x_1,x_2) = \mathbf{\widetilde \Sigma}(\mathbf{r}_1,\mathbf{r}_2) \, \delta(t_1 - t_2) + \mathbf{\Sigma}^{(e)}(x_1,x_2) \; .
\label{eq:self}
\end{eqnarray}
Note that in what follows, we indicate the static quantities with a tilde. 

\subsubsection{Static self-energy: Relativistic Hartree-Bogoliubov theory}
In this work the static part of the mass operator (\ref{eq:self}) is calculated within the Relativistic Hartree-Bogoliubov (RHB) theory. Here we only recall a few important results of this theory, for more details we refer the reader to e.g. Ref. \cite{Vretenar}. In the standard RHB framework, the energy functional is derived from the relativistic Lagrangian (\ref{eq:Lagrangian}), and is extended to account for pairing correlations in a phenomenological way. 
The mean-field approximation is then applied and consists in replacing the boson fields by their classical expectation value, i.e. only the nucleon fields are quantized. 
The "no-sea approximation" is also applied, so that antiparticle states are excluded from the expansion of the Dirac fields that read

\begin{equation}
\hat \psi(x) = \sum_\alpha \varphi_\alpha(\boldsymbol{r}) e^{-i \varepsilon_\alpha t} \hat{a}_{\alpha} \; , 
\end{equation}
where $\alpha = (n_\alpha, \pi_\alpha, j_\alpha,m_\alpha, \tau_\alpha)$ is a set of quantum numbers for the nucleon states, and $\varphi_\alpha(\boldsymbol{r})$ is a 4-component Dirac spinor with small and large components $f_\alpha$ and $g_\alpha$:
\begin{equation}
\varphi_\alpha(\boldsymbol{r}) = 
\begin{pmatrix}
f_\alpha (\boldsymbol{r}) \\
i g_\alpha (\boldsymbol{r})
\end{pmatrix}
\; .
\end{equation}

\noindent In the stationary limit, and neglecting retardation effects, a variational principle finally leads to the following system of coupled RHB equation for the nucleons and Klein-Gordon equation for the bosons $\{b=\sigma,\omega,\rho,A\}$:
\begin{numcases}{}
\int d\boldsymbol{r}_2 \widetilde{\boldsymbol{\mathcal{H}}} (\boldsymbol{r}_1,\boldsymbol{r}_2) \boldsymbol{\phi}_k^\eta(\boldsymbol{r}_2) = \eta E_k \boldsymbol{\phi}_k^\eta(\boldsymbol{r}_1)  \; , \eta= \pm  \label{eq:RHB} \\
 - \Delta \phi_b (\mathbf{r}) + U'[\phi_b (\mathbf{r})] = \mp  \langle \hat{\bar\psi}(\mathbf{r})  \Gamma_b \hat{\psi}(\mathbf{r}) \rangle \; , \label{eq:KG}
\end{numcases}
where in Eq. (\ref{eq:KG}), the $+$ ($-$) sign holds for vector (scalar) mesons.
\\
In Eq. (\ref{eq:RHB}), $\widetilde{\boldsymbol{\mathcal{H}}}$ is the RHB Hamiltonian which reads 
\begin{equation}
 \widetilde{\boldsymbol{\mathcal{H}}} = 
 \begin{pmatrix}
  h_D  - \lambda & \gamma^0 \Delta \\
  \gamma^0 \overline{\Delta} & -h_D^*  + \lambda 
 \end{pmatrix}
 \; ,
\end{equation}
where $h_D$ is the Dirac Hamiltonian
\begin{equation}
  h_D = \gamma^0 ( \gamma^i p_i + m + \widetilde{\Sigma}_{RMF}) \; ,
\end{equation}
and $\widetilde{\Sigma}_{RMF} = \sum_b \Gamma_b \phi_b$ is the relativistic mean-field (RMF) self-energy for normal systems. As we assume here time-reversal invariance, the currents vanish so that only the scalar and the time-like vector RMF self-energies remain. Finally, $\Delta$ and $\overline\Delta$ are the pairing fields which read, in the case of a local force in the particle-particle channel:
\begin{eqnarray}
\Delta(\boldsymbol{r}_1,\boldsymbol{r}_2) = V^{pp} (\boldsymbol{r}_1,\boldsymbol{r}_2) \braket{\hat{\psi}(\boldsymbol{r}_2)\hat{\psi}^T(\boldsymbol{r}_1)} \; , \\
\overline{\Delta}(\boldsymbol{r}_1,\boldsymbol{r}_2) = V^{pp} (\boldsymbol{r}_1,\boldsymbol{r}_2) \braket{\hat{\bar{\psi}}^T(\boldsymbol{r}_2)\hat{\bar{\psi}}(\boldsymbol{r}_1)} \; .
\end{eqnarray} 
The quasiparticle self-energy obtained in the RHB approximation is then static but non-local in space and reads
\begin{eqnarray}
&&\mathbf{\widetilde\Sigma}(\mathbf{r_1},\mathbf{r_2}) = \nonumber \\
&&
\begin{pmatrix}
\widetilde\Sigma_{RMF}(\mathbf{r_1}) \delta(\mathbf{r_1}-\mathbf{r_2}) & \Delta(\mathbf{r_1},\mathbf{r_2}) \\
\overline{\Delta}(\mathbf{r_1},\mathbf{r_2}) & - \widetilde\Sigma_{RMF}(\mathbf{r_1})\delta(\mathbf{r_1}-\mathbf{r_2})
\end{pmatrix}
\; . \nonumber \\
\label{eq:RHB_SE}
\end{eqnarray}

\noindent The solutions of Eq. (\ref{eq:RHB}) are the Dirac-Bogoliubov quasiparticle spinors of the form
\begin{eqnarray}
\boldsymbol{\phi}_k^+ (\boldsymbol{r}) = 
\begin{pmatrix}
\phi_k^+ (\boldsymbol{r},+) \\  
\phi_k^+ (\boldsymbol{r},-)
\end{pmatrix}
= \sum_\alpha
\begin{pmatrix}
\varphi_\alpha (\boldsymbol{r}) U_{\alpha k} \\  
\bar{\varphi}_\alpha^T (\boldsymbol{r}) V_{\alpha k} 
\end{pmatrix}
\; ,
\\
\boldsymbol{\phi}_k^- (\boldsymbol{r}) = 
\begin{pmatrix}
\phi_k^- (\boldsymbol{r},+) \\  
\phi_k^- (\boldsymbol{r},-)
\end{pmatrix}
= \sum_\alpha
\begin{pmatrix}
\varphi_\alpha (\boldsymbol{r}) V^*_{\alpha k} \\  
\bar{\varphi}_\alpha^T (\boldsymbol{r}) U^*_{\alpha k} 
\end{pmatrix}
\; .
\end{eqnarray}
The index $\eta=\pm$ serves to denote the quasiparticles with positive and negative energy $\pm E_k$ ($E_k > 0$).
Occasionally, we will refer to the solutions with negatives energies as "quasiholes".
Similarly as in (\ref{eq:general_field}) we can define corresponding generalized creation and destruction operators of quasiparticles $\hat{\boldsymbol{B}}_k = \{ \hat{B}_k^\eta\}$, with
\begin{eqnarray}
\hat{B}_k^+  &=& \hat{B}^{\dagger\,-}_k \equiv \hat{\beta}_k \\
\hat{B}_k^-   &=& \hat{B}^{\dagger\,+}_k \equiv \hat{\beta}^\dagger_k
\; ,
\end{eqnarray}
where $\hat\beta_k$ and $\hat{\beta}^\dagger_k$ are the usual Bogoliubov quasiparticle operators \cite{RingSchuck}.
They are related to the field operators through
\begin{equation}
\hat \Psi (\boldsymbol{r},\chi) = \sum_{k,\eta} \phi_k^\eta (\boldsymbol{r}, \chi) \hat{B}^\eta_k \; .
\end{equation}

\begin{figure}
\centering
\resizebox{0.07\textwidth}{!}{%
\includegraphics{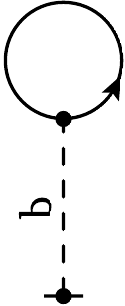}}
\caption{Static RHB self-energy. The full line represents the propagator of a Bogoliubov quasiparticle in the $(k,\eta)$ basis. The dashed line denotes the propagator of a boson $b=\sigma,\omega^\mu,\protect\overrightarrow{\rho}^\mu, A^\mu $.}
\label{fig:SE_RHB}
\end{figure}

\begin{figure}
\centering
\resizebox{0.4\textwidth}{!}{%
\includegraphics{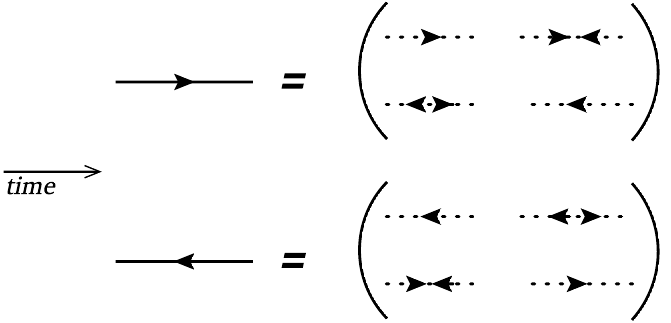}}
\caption{Mean-field quasiparticle propagator $G^{ \eta_1, \eta_2}_{k_1, k_2} (\varepsilon) \propto  \delta_{ \eta_2,\eta_1} \delta_{k_1, k_2}$, expanded in the doubled coordinate space $(x,\chi)$. We show the case $\eta_1 = +$ (top), and $\eta_1 = -$ (bottom). The dotted lines represent the normal and anomalous Gorkov propagators $G(x_1,\chi_1;x_2,\chi_2)$.}
\label{fig:GF_RHB}
\end{figure}

\noindent In the basis $\{k,\eta\}$ of these operators, the mean-field propagator solution of $\mathbf{\widetilde{G}}= \mathbf{G}^0 + \mathbf{G}^0 \mathbf{\widetilde{\Sigma}}\mathbf{\widetilde{G}}$ has a simple diagonal form 
\begin{equation}
 \mathbf{\widetilde G}_{k_1, k_2} (t_{12}) = 
 \begin{pmatrix}
  \widetilde G_{k_1, k_2}^{+,+}(t_{12})  & 0 \\
  0 & \widetilde G_{k_1,k_2}^{-,-}(t_{12}) 
 \end{pmatrix}
 \; ,
\end{equation}
with components
\begin{equation}
 \widetilde G_{k_1, k_2}^{\eta_1,\eta_2}(t_{12}) = -i \eta_1 \delta_{k_1,k_2} \delta_{\eta_1,\eta_2}  \theta\left(\eta_1 (t_{12}) \right) e^{-i \eta_1 E_{k_1} t_{12}} \; ,
 \label{eq:RHB_propag_time}
\end{equation}
where $t_{12} \equiv t_1 - t_2$. Equivalently, in the energy representation
\begin{equation}
\widetilde{G}^{\eta_1 \eta_2}_{k_1 k_2} (\varepsilon)= \frac{\delta_{k_1 k_2} \delta_{\eta_1 \eta_2} }{\varepsilon - \eta_1 E_{k_1} + i \eta_1 \delta} \; , \; \delta\rightarrow 0^+ \; .
\label{eq:RHB_propag}
\end{equation}
In this basis, the propagator of a quasiparticle with definite energy can then be represented by a single line, propagating inward or backward in time. The self-energy (\ref{eq:RHB_SE}) can therefore be drawn as in Fig. \ref{fig:SE_RHB}. When expressed in the extended coordinate space $(x,\chi)$ this propagator has a non-diagonal structure and contains the usual normal and anomalous Green's functions. This is schematically represented in Fig. \ref{fig:GF_RHB}.

\subsubsection{Energy-dependent self-energy: quasiparticle-vibration coupling}
\begin{figure}
\centering
\resizebox{0.4\textwidth}{!}{%
\includegraphics{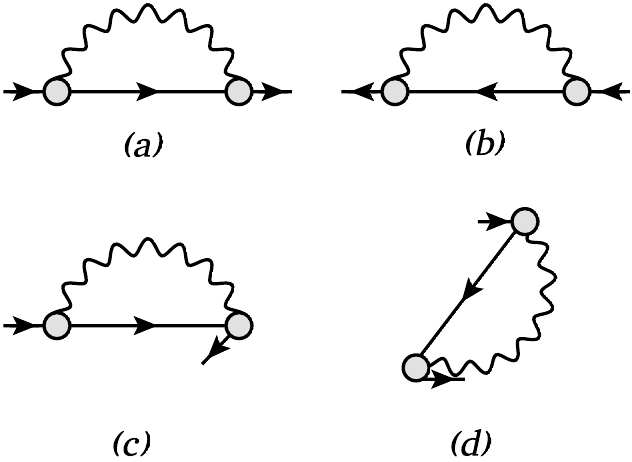}}
\caption{Examples of energy-dependent quasiparticle self-energy components $\Sigma^{(e)\,\eta_1 \eta_2}_{k_1 k_2}(\varepsilon)$. The wavy line represents a collective phonon, while the solid straight line represents a quasiparticle propagator (\ref{eq:RHB_propag}) in the RHB basis. The grey dots are the quasiparticle-vibration coupling vertices $\gamma$. The external lines are not included in the self-energy. The forward-going diagrams (a) and (b) represent the case $\eta_1= \eta_2 = \eta_\mu = +$ and $\eta_1= \eta_2 = \eta_\mu = -$ respectively. The backward-going diagrams (c) and (d) represent the case $\eta_1=\eta_\mu=+=-\eta_2$ and $\eta_1 = \eta_2 = + = -\eta_\mu$ respectively. Note that in the present applications, the backward-going diagrams will not be implemented.}
\label{fig:dyn_SE}
\end{figure}

When the dynamics is taken into account, the oscillations of the nuclear mean-field lead to surface vibrations, also called phonons, that can couple to the motion of single quasiparticles. In this work, the energy-dependent part of the self-energy (\ref{eq:self}) is therefore obtained within the Quasiparticle-Vibration Coupling (QPVC) model. In the RHB quasiparticle basis, the components of $\mathbf{\Sigma}^{(e)}$ take the following expression in the energy representation \cite{Litvinova2008}:
\begin{eqnarray}
\Sigma^{(e)\,\eta_1 \eta_2}_{k_1 k_2}(\varepsilon) = &&\sum_{\eta=\pm 1} \sum_{\eta_\mu=\pm 1} \sum_{k,\mu} \frac{\delta_{\eta_\mu,\eta} \gamma_{\mu ;k_1 k}^{\eta_\mu ; \eta_1 \eta}  \gamma_{\mu ;k_2 k}^{\eta_\mu ; \eta_2 \eta *}}{\varepsilon - \eta E_k - \eta_\mu (\Omega_\mu - i\delta)} \; , \nonumber \\
&& \delta \rightarrow 0 \; ,
\label{eq:dyn_SE}
\end{eqnarray}
where the index $\mu$ labels the collective phonons of frequency $\Omega_\mu$ that couple to the quasiparticle states through the vertices 
\begin{equation}
\gamma_{\mu ;k_1 k_2}^{\eta_\mu ; \eta_1 \eta_2} = \delta_{\eta_\mu,+1} \gamma_{\mu ;k_1 k_2}^{\; \eta_1 \eta_2} + \delta_{\eta_\mu,-1} \gamma_{\mu ;k_2 k_1}^{\; \eta_2 \eta_1*} \; .
\end{equation}
Depending on $\eta_1$ and $\eta_2$, the sign $\eta = \eta_\mu$ corresponds to forward and backward going diagrams, as seen in Fig. \ref{fig:dyn_SE} which displays some possible self-energy terms. As will be emphasized later, the backward-going diagrams, such as (c) and (d), account for configurations that are more complex than 1qp$\otimes$phonon, and will be neglected in this work.

\noindent In the present applications the quasiparticle-vibration vertices are obtained from the solutions of the relativistic QRPA as
\begin{equation}
\gamma_{\mu ;k_1 k_2}^{\; \eta_1 \eta_2} = \sum_{k_3 k_4} \sum_{\eta_3 \eta_4} \widetilde{V}_{k_1 k_4 , k_2 k_3}^{\eta_1 \eta_4 \eta_2 \eta_3} \mathcal{R}_{\mu;k_3 k_4}^{\; \eta_3 \eta_4} \; ,
\end{equation}
where 
\begin{equation}
\widetilde{V}_{k_1 k_4 , k_2 k_3}^{\eta_1 \eta_4 \eta_2 \eta_3} = \frac{\delta \widetilde{\Sigma}_{k_4 k_3}^{ \eta_4 \eta_3}}{\delta \mathcal{R}_{k_2 k_1}^{\eta_2 \eta_1}}
\label{eq:static_int}
\end{equation}
is the static interaction defined as the functional derivative of the RHB self-energy (\ref{eq:RHB_SE}) with respect to the generalized density matrix $\mathcal{R}$ of the RHB ground state, and the set of $\mathcal{R}_{\mu}$ are the one-body transition densities obtained within the RQRPA approximation. For more details, see \cite{Litvinova2008}.

\subsection{Response of nuclei to an isospin-transfer external field}
\begin{figure*}
\centering
\includegraphics{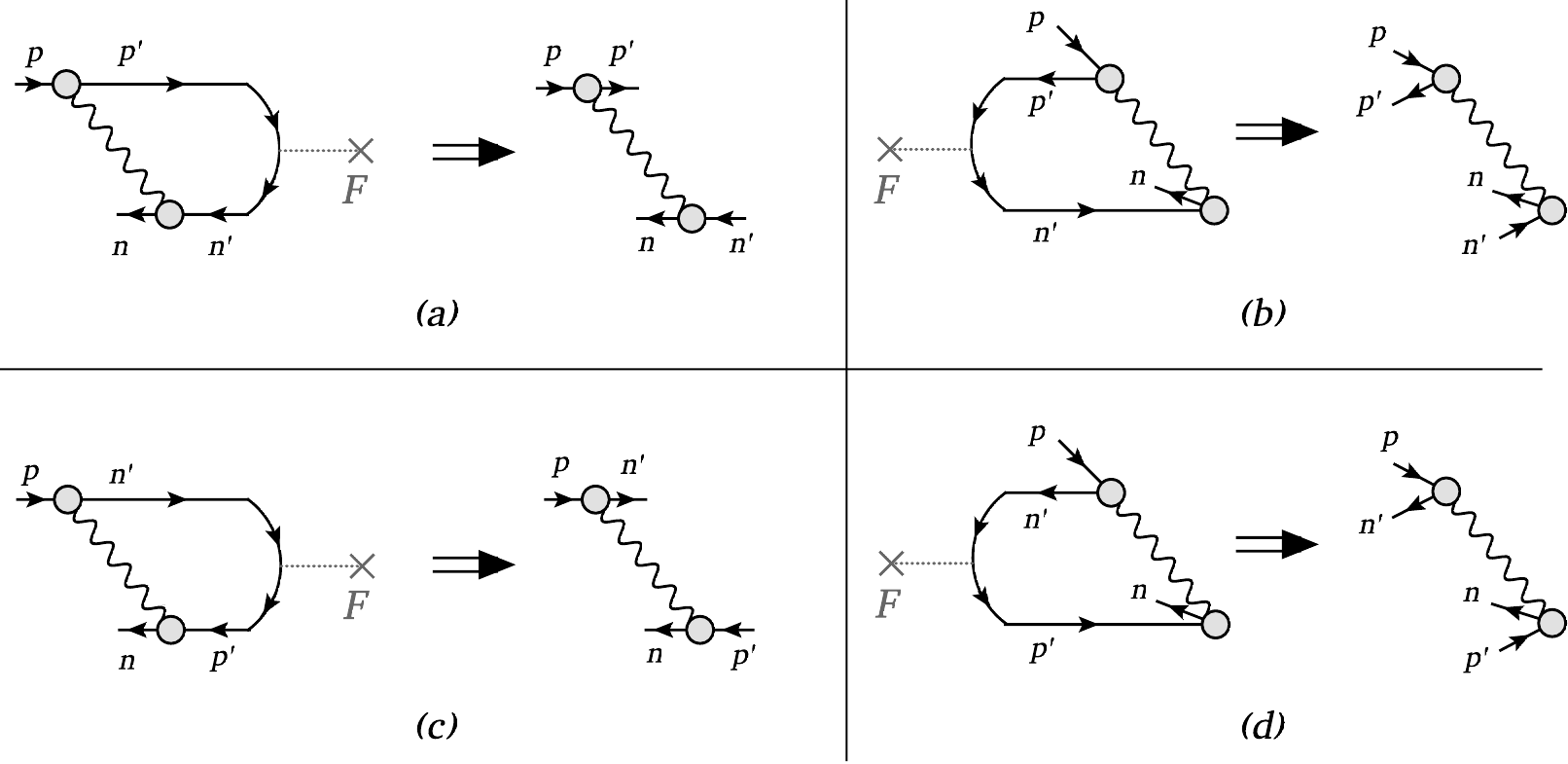}
\caption{Examples of self-energy components $\Sigma_F^{(e)\eta_p; \eta_n}(p;n)$ in the presence of a charge-exchange external field $F$ of quasiparticle-quasihole nature, and corresponding effective interaction terms $V^{(e)}$. The diagrams (a) and (b) are obtained with isoscalar phonons, while the diagrams (c) and (d) are generated by isovector phonons. The backward-going diagrams (b) and (d) are responsible for ground-state correlations. In the present work, only diagrams of type (a) are implemented.}
\label{fig:eff_int}
\end{figure*}

When the nucleus is subjected to a one-body external perturbation $F$, the linear variation of the single-quasiparticle propagator leads to the response function
\begin{eqnarray}
\boldsymbol{R}(14,23) &\equiv& \frac{\delta \boldsymbol{G}_F (1,2)}{ \delta \boldsymbol{F}(3,4)}_{|F=0} \label{eq:resp1} \\
               &=& - \boldsymbol{G}(14,23) + \boldsymbol{G}(1,2) \boldsymbol{G}(4,3) \label{eq:resp2} \; ,
\end{eqnarray}
where we have introduced the index $1 \equiv (k_1,t_1)$. $\boldsymbol{G}_F$ denotes the propagator of a single quasiparticle in the nucleus in the presence of the source field $F$ (i.e. when the nucleus is governed by $H'' = H'+F $). On line (\ref{eq:resp2}), $\boldsymbol{G}(1,2)$ and $\boldsymbol{G}(14,23)$, without subscript $F$, denote equilibrium one- and two-body Green's functions respectively, as defined in Eqs. (\ref{eq:GF1}), and coincide with Green's functions in the source field in the limit $F\rightarrow 0$.
From Eq. (\ref{eq:resp2}), we see that the response function $\boldsymbol{R}$ describes the propagation of two quasiparticles in the nuclear medium.
\\
Taking the functional derivative of the Dyson equation for $G_F(1,2)$ with respect to the external field $F$ leads to the Bethe-Salpeter equation for the response function \cite{Speth,Tselyaev}:
\begin{eqnarray}
\boldsymbol{R}(14,23) &=& \boldsymbol{G}(1,3) \boldsymbol{G}(4,2) \nonumber \\ 
                 &-&i \int d5... d8 \boldsymbol{G}(1,5) \boldsymbol{G}(6,2) \boldsymbol{V}(58,67) \boldsymbol{R}(74,83) \; , \nonumber \\
\label{eq:BSE}
\end{eqnarray}
where $\boldsymbol{V}$ is the effective interaction determined by the functional derivative of the self-energy with respect to the single quasiparticle propagator:
\begin{eqnarray}
\boldsymbol{V}(58,67) &\equiv& i \frac{\delta \boldsymbol{\Sigma}_F(5,6)}{\delta \boldsymbol{G}_F(7,8)}_{|F=0} \nonumber \\
                &=&    i \frac{\delta \widetilde{\boldsymbol{\Sigma}}_F(5,6)}{\delta \boldsymbol{G}_F(7,8)}_{|F=0}  + i \frac{\delta \boldsymbol{\Sigma}^{(e)}_F(5,6)}{\delta \boldsymbol{G}_F(7,8)}_{|F=0}     \nonumber \\
                &=&    \widetilde{\boldsymbol{V}}(58,67) + \boldsymbol{V}^{(e)}(58,67)    \; .
\label{eq:eff-int}
\end{eqnarray}
We note that in this work, both the static and dynamical self-energies $\boldsymbol{\widetilde{\Sigma}}$ and $\boldsymbol{{\Sigma}^e}$ are actually constructed with the mean-field propagator $\boldsymbol{\widetilde{G}}$ instead of the full one $\boldsymbol{G}$. In principle, one could imagine building an iterative procedure, where the new propagator would be used to calculate the self-energy, which in turn would lead to a renormalized Green's function, and so on.
In Fig. \ref{fig:eff_int} we show possible components of the self-energy $\boldsymbol{\Sigma}_{F}^{(e)}(p,n)$ in an isospin-transfer external field $F$ of quasiparticle-quasihole nature, as well as the corresponding effective interaction terms $\boldsymbol{V}^{(e)}$. The index $\tau=(p,n)$ denotes proton or neutron states ($k_\tau,t_\tau$). Diagrams (a) and (b) show the energy-independent interaction generated by the exchange of isoscalar collective vibrations, while diagrams (c) and (d) are produced by isovector phonons. In this work, the coupling of quasiparticles to isovector vibrations are not considered. And again, due to the level of approximation considered in this work, the backward-going diagrams leading to ground-state correlations due to phonon-coupling are also neglected. \\ \\

\noindent The BSE (\ref{eq:BSE}), that is expressed in terms of the full quasiparticle propagator $\mathbf{G}$, can then be recasted in terms of the free RHB propagator $\mathbf{\widetilde G}$ given by Eq. (\ref{eq:RHB_propag_time}), using the Dyson equation
\begin{equation}
\boldsymbol{\widetilde{G}}^{-1}(1,2) = \boldsymbol{G}^{-1}(1,2) + \boldsymbol{\Sigma}^{(e)}(1,2) \; .
\end{equation}
Finally, similarly to the neutral channel \cite{Litvinova2008}, the Relativistic Quasiparticle Time-Blocking Approximation (RQTBA) \cite{TBA} is applied. This approximation makes it possible to decouple the energy integrations in the BSE by introducing a time-projection operator which blocks the configurations beyond 4 quasiparticle excitations. In particular, the backward-going diagrams in the self-energy in Fig. (\ref{fig:dyn_SE}), as well as in the dynamical effective interaction in Fig. \ref{fig:eff_int}, are then neglected\footnote{Work to include higher-order configurations through an iterative procedure is already underway in the neutral channel \cite{Litvinova2015}.}. Ground-state correlations are therefore introduced via the static interaction $\boldsymbol{\widetilde V}$ only.
\\
In the quasiparticle-quasihole channel, i.e. in the time limit ($t_1 = t_2$, $t_3=t_4$), and after a Fourier transform to the energy domain, we obtain the following BSE for the response function in the proton-neutron channel:

\begin{eqnarray}
&& R^{\eta_1 \eta_4 , \eta_2 \eta_3}_{k_{1_p} k_{4_n}, k_{2_n} k_{3_p}} (\omega) = \widetilde{R}^{(0) \eta_1 \eta_4 , \eta_2 \eta_3}_{k_{1_p} k_{4_n}, k_{2_n} k_{3_p}} (\omega) \nonumber \\
&& \hspace{1.5cm}   + \sum_{k_{5_p} k_{6_n} k_{7_p} k_{8_n}} \sum_{\eta_5 \eta_6 \eta_7 \eta_8} \widetilde{R}^{(0) \eta_1 \eta_6 , \eta_2 \eta_5}_{k_{1_p} k_{6_n}, k_{2_n} k_{5_p}} (\omega) \nonumber \\
 && \hspace{1.5cm}  \times \; \bar{W}^{\eta_5 \eta_8 , \eta_6 \eta_7}_{k_{5_p} k_{8_n}, k_{6_n} k_{7_p}} (\omega) R^{\eta_7 \eta_4 , \eta_8 \eta_3}_{k_{7p} k_{4_n}, k_{8_n} k_{3_p}} (\omega) \; ,  \nonumber \\
 \label{eq:BSE_pn}
\end{eqnarray}
 where the indices $k_{i_p}$ and $k_{i_n}$ denote proton and neutron single-particle states respectively, and 
\begin{eqnarray}
&&R^{\eta_1 \eta_4 , \eta_2 \eta_3}_{k_{1_p} k_{4_n}, k_{2_n} k_{3_p}} (\omega) = \delta_{\eta_1, -\eta_2} \delta_{\eta_3, -\eta_4} \times (-i) \nonumber  \\
&&\times \int dt_1 ... dt_4 \delta(t_1-t_2) \delta(t_3-t_4) \delta(t_4) e^{i\omega (t_1 - t_3)} R(14,23) \nonumber \; .
\end{eqnarray}
In Eq. (\ref{eq:BSE_pn}), the free response $R^{(0)}(\omega)$ reads
\begin{eqnarray}
R^{(0)\eta_1 \eta_4 , \eta_2 \eta_3}_{k_{1_p} k_{4_n}, k_{2_n} k_{3_p}} (\omega) = \frac{  \delta_{\eta_1,\eta_3} \delta_{\eta_1, -\eta_2} \delta_{\eta_3, -\eta_4} \delta_{k_{1_p}, k_{3_p}} \delta_{k_{2_n}, k_{4_n}}  }{\eta_1 (\omega-\lambda_{pn}) - E_{k_{1_p}} - E_{k_{2_n}}} \; ,\nonumber \\
\end{eqnarray}
where $\lambda_{pn} = \lambda_{p} - \lambda_{n}$ is the difference between proton and neutron chemical potentials.
The effective interaction $\bar{W}$ is the sum of the static interaction and a dynamical one:
\begin{equation}
\bar{W}^{\eta_5 \eta_8 , \eta_6 \eta_7}_{k_{5_p} k_{8_n}, k_{6_n} k_{7_p}} (\omega) = \widetilde{V}^{\eta_5 \eta_8 , \eta_6 \eta_7}_{k_{5_p} k_{8_n}, k_{6_n} k_{7_p}}  + \Phi^{\eta_5 \eta_8 , \eta_6 \eta_7}_{k_{5_p} k_{8_n}, k_{6_n} k_{7_p}} (\omega) \; .
\label{eq:eff_int}
\end{equation}

\noindent In the proton-neutron channel, only the isovector part $\widetilde{V}^{(IV)}$ of the static interaction $\widetilde{V}$ contributes to the BSE (\ref{eq:BSE_pn}). This part describes pion and rho-meson exchange:

\begin{eqnarray}
\widetilde{V}^{(IV)} (\mathbf{r_1},\mathbf{r_2}) &=& \widetilde V_\rho (\mathbf{r_1},\mathbf{r_2}) + \widetilde V_\pi (\mathbf{r_1},\mathbf{r_2}) + \widetilde V_{\delta_{\pi}} (\mathbf{r_1},\mathbf{r_2}) \nonumber \\
                     &=& \bigl( \sum_{b=\rho,\pi} \Gamma_1^b D_b  (\mathbf{r_1} - \mathbf{r_2}) \Gamma^b_2 \bigr) + \widetilde V_{\delta_{\pi}} (\mathbf{r_1},\mathbf{r_2}) \nonumber \\ 
                     &=& g_\rho^2 \overrightarrow{\tau_1} \overrightarrow{\tau_2} (\gamma^\mu)_1 (\gamma_\mu)_2 D_\rho(\mathbf{r_1}-\mathbf{r_2}) \nonumber \\
                     && - \left( \frac{f_\pi}{m_\pi} \right)^2 \overrightarrow{\tau_1} \overrightarrow{\tau_2} (\boldsymbol{\hat\sigma_1} \boldsymbol{\nabla_1}) (\boldsymbol{\hat\sigma_2} \boldsymbol{\nabla_2}) D_\pi(\mathbf{r_1}-\mathbf{r_2}) \nonumber \\
                     && + g' \left( \frac{f_\pi}{m_\pi} \right)^2 \overrightarrow{\tau_1} \overrightarrow{\tau_2}\boldsymbol{\hat\sigma_1} \boldsymbol{\hat\sigma_2} \delta(\mathbf{r_1}-\mathbf{r_2}) \; ,
 \label{eq:iso_stat_int}                    
\end{eqnarray}
where $D_b$ are the meson propagators 
\begin{equation}
D_b (\mathbf{r_1} - \mathbf{r_2})  = \frac{1}{4\pi} \frac{e^{-m_b |\mathbf{r_1} - \mathbf{r_2}|}}{ |\mathbf{r_1} - \mathbf{r_2}|} \; ,
\end{equation}
and $\boldsymbol{\hat\sigma}$ is the generalized Pauli matrix \cite{Paar}
\begin{equation}
\boldsymbol{\hat\sigma} = 
\begin{pmatrix}
\boldsymbol{\sigma} & 0 \\
0 & \boldsymbol{\sigma} 
\end{pmatrix}
\; .
\end{equation}

\begin{figure*}
\centering
\includegraphics{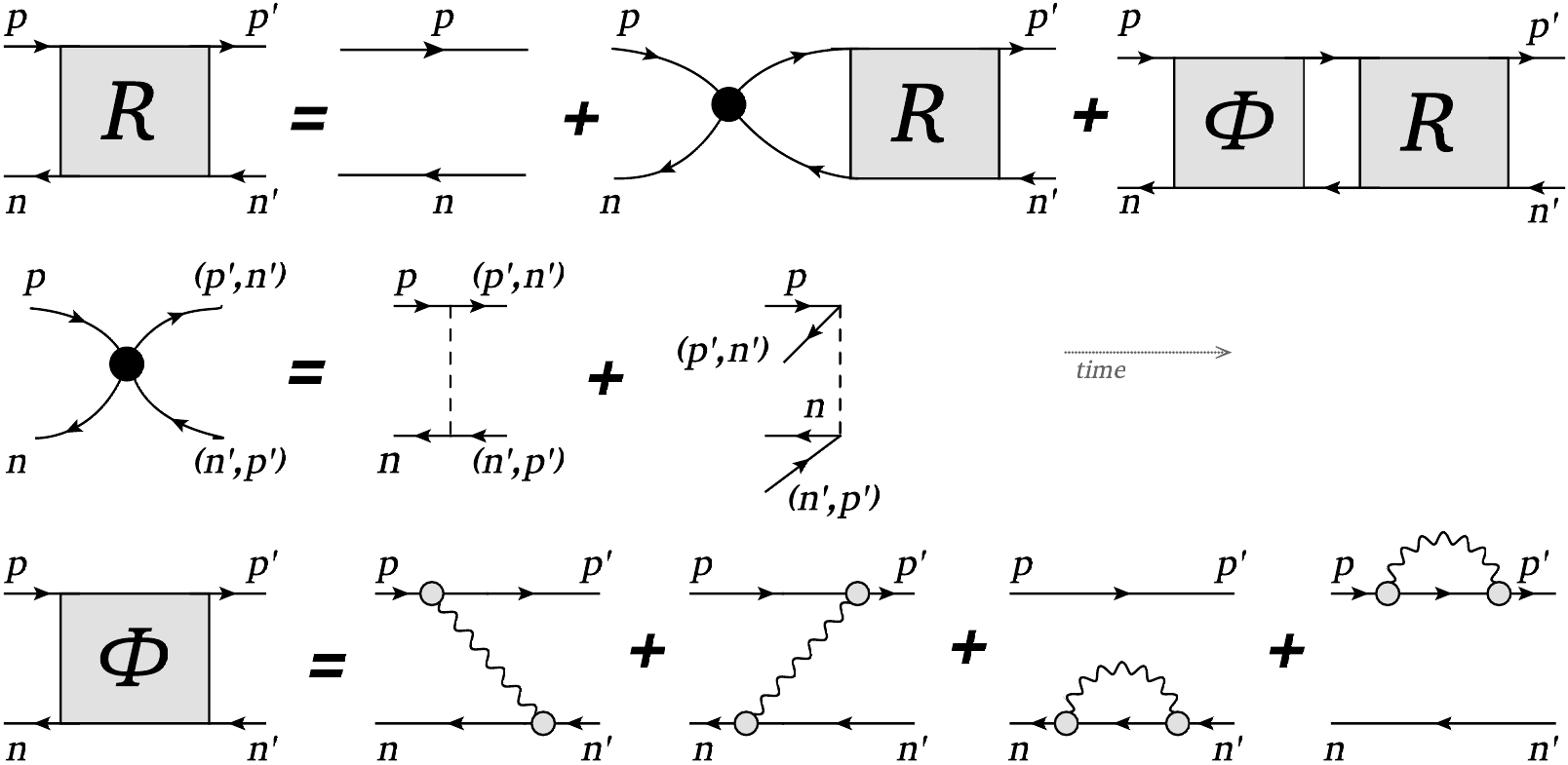}
\caption{Bethe Salpeter equation (\ref{eq:BSE_pn}) in the proton-neutron channel in the RQTBA approximation.}
\label{fig:BSE_pn}
\end{figure*}

\noindent The Landau-Migdal term $V_{\delta_\pi}$ accounts for short-range correlations \cite{Bouyssy}. The strength of these correlations is governed by the parameter $g'$ with the  fixed bare value equal to $\frac{1}{3}$. This value is used in the calculations of Gamow-Teller and spin-dipole resonances within the relativistic random phase approximation based on the relativistic Hartree-Fock mean field \cite{Liang}. However, in the absence of the explicit Fock term in the mean field, the pionic contribution vanishes and, as a result, the consistency between the mean field and the effective interaction of Eq. (\ref{eq:iso_stat_int}) becomes elusive. An additional constraint on the parameter $g'$ is therefore needed and, following the literature \cite{Paar}, we adopt the value $g'=0.6$, i.e. about twice the bare value of $\frac{1}{3}$. This value allows one to reproduce the position of the Gamow-Teller resonance in $^{208}$Pb, and is kept identical for all other nuclei. A fully self-consistent pn-RQTBA description of spin-isospin resonances would imply the explicit inclusion of the Fock term, the use of the bare value for the $g'$ constant, and restoration of the ''subtraction'' procedure (see below). This program is, however, beyond the scope of the present article and will be considered in the future work. 
\\
Considering only the static part (\ref{eq:iso_stat_int}) of the effective interaction (\ref{eq:eff-int}), the BSE (\ref{eq:BSE_pn}) reduces to the proton-neutron Relativistic QRPA (pn-RQRPA) which only accounts for one-quasiparticle excitations.
\\
\\
Finally, $\boldsymbol{\Phi}(\omega)$ is an energy-dependent interaction which describes the coupling between quasiparticles and collective phonons and reads in the time-blocking approximation
\begin{eqnarray}
&&\Phi^{\eta_1 \eta_4 , \eta_2 \eta_3}_{k_{1_p} k_{4_n}, k_{2_n} k_{3_p}} (\omega) =  \delta_{\eta_1,\eta_3} \delta_{\eta_1, -\eta_2} \delta_{\eta_3, -\eta_4} \times \sum_{\mu \xi} \delta_{\xi,\eta_1} \nonumber \\
\times \Bigl[ && \delta_{k_{1_p} k_{3_p}} \sum_{k_{6_n}} \frac{\gamma_{\mu ;k_{6_n} k_{2_n}}^{\eta_1; -\xi -\xi} \gamma_{\mu ;k_{6_n} k_{4_n}}^{\eta_1; -\xi -\xi *}}{\eta_1 (\omega - \lambda_{pn}) - E_{k_{1_p}} -E_{k_{6_n}} - \Omega_\mu } \nonumber \\
&&+ \delta_{k_{2_n} k_{4_n}} \sum_{k_{5_p}} \frac{\gamma_{\mu ;k_{1_p} k_{5_p}}^{\eta_1; \xi \xi} \gamma_{\mu ;k_{3_p} k_{5_p}}^{\eta_1; \xi \xi *} }{\eta_1 (\omega - \lambda_{pn}) - E_{k_{5_p}} -E_{k_{2_n}} - \Omega_\mu} \nonumber \\
&&- \frac{\gamma_{\mu ;k_{1_p} k_{3_p}}^{\eta_1; \xi \xi} \gamma_{\mu ;k_{2_n} k_{4_n}}^{\eta_1; -\xi -\xi *} }{\eta_1 (\omega - \lambda_{pn}) - E_{k_{3_p}} -E_{k_{2_n}} - \Omega_\mu} \nonumber \\
&&- \frac{\gamma_{\mu ;k_{3_p} k_{1_p}}^{\eta_1; \xi \xi *} \gamma_{\mu ;k_{4_n} k_{2_n}}^{\eta_1; -\xi -\xi } }{\eta_1 (\omega - \lambda_{pn}) - E_{k_{1_p}} -E_{k_{4_n}} - \Omega_\mu} \Bigr] \;. \nonumber \\
\label{eq:Phi}
\end{eqnarray}
The first two terms on the right-hand side of Eq. (\ref{eq:Phi}) describe dynamical self-energy insertions whereas the two last ones describe the exchange of collective phonons between two quasiparticle states. By accounting for configurations of the type 1qp$\otimes$phonon, i.e. 2qp excitations, this term is responsible for the fragmentation of the many-body states.
The final BSE (\ref{eq:BSE_pn}) in the proton-neutron channel is represented in Fig. \ref{fig:BSE_pn}. 
\\ 
\\
As the present approach is based on an energy functional with parameters adjusted so to reproduce properties of nuclei at the RMF level, usually, in the isospin-like (neutral) channel, a "subtraction procedure" needs to be applied \cite{Tselyaev}. This technique consists in subtracting from the phonon-coupling interaction (\ref{eq:Phi}) its value at zero frequency $\Phi(\omega = 0)$. 
In the neutral channel, where the same mesons contribute to the ground state and to the description of excited states, this subtraction amounts to removing the phonon-coupling effects that are implicitly contained in the static interaction $\widetilde{V}$ through the fitting of the meson masses and vertices at the mean-field stage.
In the proton-neutron channel, considered in the present work, only the pion and $\rho$-meson contribute to the static interaction in the Bethe-Salpeter equation. While the rho-meson is present in the ground state, its relative contribution to the total isovector interaction $\widetilde{V}^{(IV)}$ has been found to be negligible in the charge-exchange channel \cite{Liang2012}. The pion, on the other hand, provides the main contribution and is absent in the RMF calculation as it would lead to parity breaking at the Hartree level. It is, therefore, considered here with the (unadjusted) free-space coupling constant ($\frac{f_\pi^2}{4\pi} =  0.08$). For these reasons we expect the double counting of quasiparticle-vibration coupling effects to be negligibly small in the charge-exchange channel and, therefore, do not apply the subtraction procedure. However, in a fully self-consistent approach based on the Hartree-Fock mean field the subtraction should be restored to avoid double counting, as it is done in RQTBA for non-isospin-flip excitations.
\\

Finally, once the response has been calculated, the spectrum of the nucleus under study is obtained by computing the strength function $S(E)$
\begin{equation}
S(E)=\frac{-1}{\pi} \lim_{\Delta \rightarrow 0} \mbox{Im} \Pi(\omega=E+i\Delta) \; ,
\label{eq:strength}
\end{equation}
proportional to the imaginary part of the polarizability $\Pi(\omega)$, which is equal to the contraction of the response function $\boldsymbol{R}(\omega)$ in the external field under consideration:
\begin{equation}
 \Pi(\omega)=\langle \boldsymbol{F}^\dagger \boldsymbol{R}(\omega) \boldsymbol{F} \rangle \; .
 \label{eq:pol}
 \end{equation}

\section{Applications to Gamow-Teller resonances in neutron-rich Nickel isotopes}
\label{sec:appli}

\subsection{Numerical scheme}
The calculations are performed according to the following scheme:
\begin{itemize}
 \item As a first step, the basis of mean-field quasiparticles $(k,\eta)$ that serves to solve the BSE (\ref{eq:BSE_pn}) is obtained by solving the RHB equation (\ref{eq:RHB}) for the nucleons together with the Klein-Gordon equation (\ref{eq:KG}) for the mesons and the photon. In the present applications we use the NL3 \cite{NL3} parametrization of the meson-exchange interaction, as well as a monopole-monopole pairing force \cite{Litvinova2008} for which the RHB equation is equivalent to the RH+BCS problem. The isospin-like pairing correlations are included in a smooth pairing window of 20 MeV around the Fermi energy. This means that they are absent from the Dirac sea of antiparticle states. The parameter of the pairing interaction is adjusted so as to reproduce the gap at the Fermi level given by the three-point formula:
 \begin{eqnarray}
 \Delta_\tau^{(3)} (A,N_\tau) &=& - \frac{(-)^{N_\tau}}{2} \Bigl( B(A-1,N_\tau - 1) \nonumber \\
 &+& B(A+1,N_\tau+1) - 2 B(A,N_\tau) \Bigr) \; , \nonumber
 \end{eqnarray}
 where $N_\tau=Z$ or $N$, for $\tau=p$ or $n$ respectively, and $B(A,N_\tau)$ is the experimental binding energy of the nucleus with A nucleons and $N_\tau$ protons or neutrons.
\item The phonons $\mu$ entering the QPVC amplitude (\ref{eq:Phi}) and their coupling vertices $\gamma_\mu$ are then obtained by solving the RQRPA equations in the resulting RH+BCS quasiparticle basis. In this work we select a set of isoscalar phonons with natural parities which are truncated according to their angular momentum and excitation energy. The values of the selected angular momenta $J_{\mu}^\pi$, and phonon energy cut-off $\Omega_{max}$ are investigated in the following paragraph \ref{cv_study}. Moreover, only the phonons realizing at least $5\%$ of the highest transition probability for a given $J_{\mu}^\pi$ are taken into account.
 \item The BSE (\ref{eq:BSE_pn}) is solved for the proton-neutron response function, with quantum numbers $J^\pi=1^+$ in the case of Gamow-Teller transitions. The quasiparticle-quasihole pairs entering Eq. (\ref{eq:BSE_pn}) are truncated according to their excitation energy, at 100 MeV in the Fermi sector and -1800 MeV for antiparticle-quasihole excitations. Because of computational limitations, the pairs participating to the quasiparticle-phonon coupling are however limited to an energy window equal to 30 MeV. Outside of this window the BSE (\ref{eq:BSE_pn}) thus reduces to the pn-RQRPA problem. 
 \item Finally, the Gamow-Teller strength distribution $S(E)$ is obtained via Eqs. (\ref{eq:strength}) and (\ref{eq:pol}), with external field operator $\hat F = \sum_i \boldsymbol{\hat\sigma}^{(i)} \tau_-^{(i)}$.
\end{itemize}

\subsection{Results}
We now apply the numerical procedure described above to the calculation of Gamow-Teller strength and $\beta$-decay half-lives of neutron-rich even-even Nickel isotopes from $^{68}$Ni to $^{78}$Ni, which are of special interest for astrophysical studies.

\subsubsection{Strength distributions}

\paragraph{Preliminary convergence study} \label{cv_study}


\begin{figure*}
\centering
\resizebox{0.9\textwidth}{!}
{
  \includegraphics{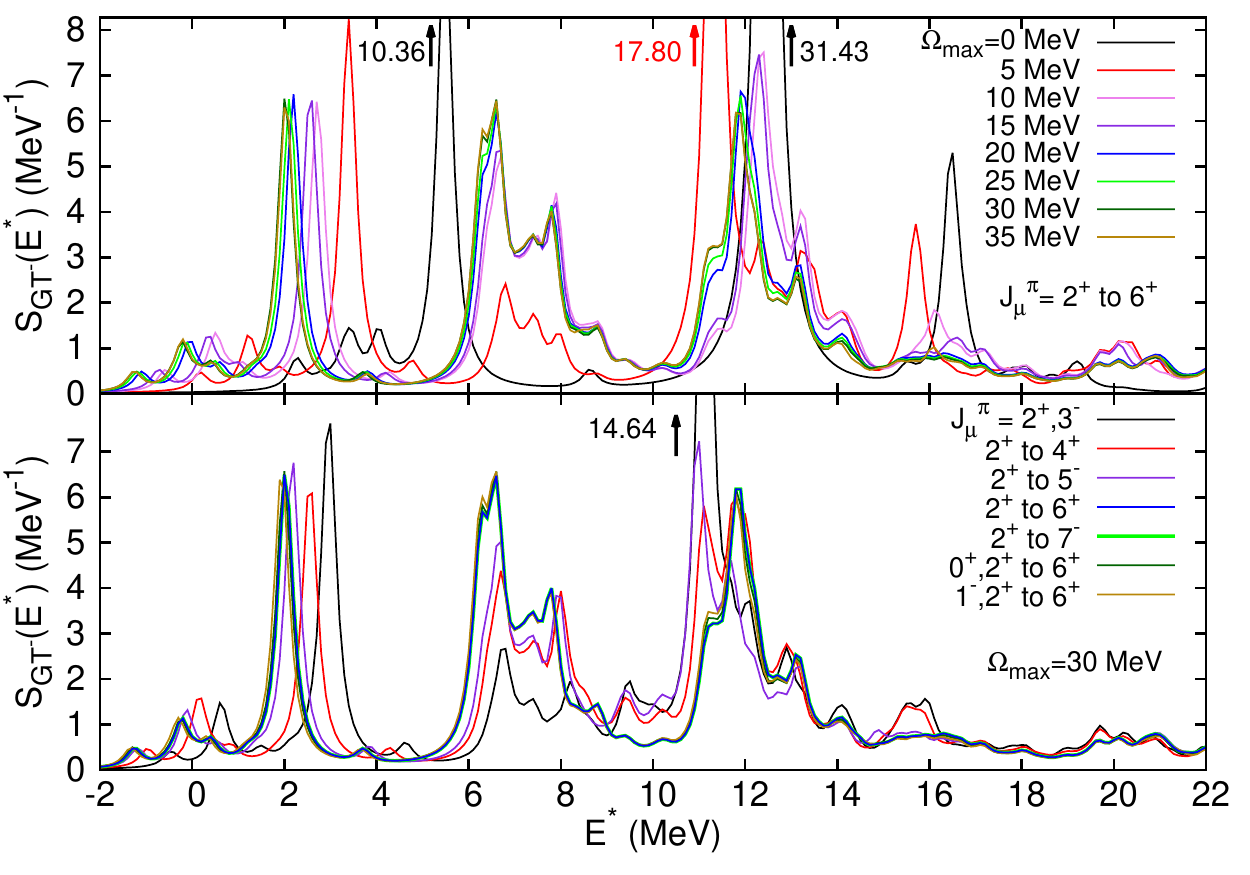}
}
\caption{(Color online) Gamow-Teller strength distributions $^{68}$Ni $\rightarrow$ $^{68}$Cu for different truncations of the phonon spectrum. Top: truncation according to the excitation energy of the phonons with quantum numbers $J^\pi_\mu = 2^+,3^-,4^+,5^-,6^+$. Bottom: truncation according to the angular momentum of the phonons with energy up to $\Omega_{max}= 30$ MeV.}
\label{fig:68Ni_Ephon_Jphon}      
\end{figure*}


In order to determine the phonon spectrum that will be included in the quasiparticle-vibration coupling, we first investigate the sensitivity and convergence of the Gamow-Teller strength in $^{68}$Ni according to the phonon angular momenta and energies. The calculated distributions are obtained by using a smearing parameter $\Delta = 200$ keV in Eq. (\ref{eq:strength}). This parameter represents an additional width attributed to each excitation, and somehow simulates the effect of higher-order configurations and continuum that are not explicitly considered in the present approach. The results are shown in Fig. \ref{fig:68Ni_Ephon_Jphon}. 
\\
The top panel displays the strengths obtained when including vibrations with quantum numbers $J^\pi_{\mu} = 2^+$ to $6^+$, while varying the energy cut-off $\Omega_{max}$. The black line, obtained with $\Omega_{max}=0$ MeV, corresponds to the pn-RQRPA result. We note that the phonons with energies up to $10$ MeV introduce the major part of the fragmentation of the strength. Increasing the phonon truncation energy to $15$ MeV only slightly modifies the results, by shifting the distribution by $\sim 100$ KeV towards lower energies. This is due to the fact that a very small number of phonons are present in the energy window $10$-$15$ MeV, in particular, no $2^+$ or $4^-$ vibrations are found in this interval. When increasing $\Omega_{max}$ up to $20$ MeV, the giant resonance region is now included in the quasiparticle-vibration coupling, and the strength is more importantly modified. Increasing again the phonon truncation energy, we note that the strength distribution tends to converge and becomes stable for $\Omega_{max}=30$ MeV. 
\\
The bottom panel of Fig. \ref{fig:68Ni_Ephon_Jphon}  displays the Gamow-Teller strength distributions obtained with an energy cut-off $\Omega_{max}$ $=30$ MeV, while including different phonon angular momenta. We note that vibrations with $J^\pi_{\mu} = 2^+$, $3^-$, $4^+$, $5^-$, $6^+$ are the most important for the description of the strength. Phonons with $J^\pi_{\mu}=1^-$ have a very small influence on the results, and the inclusion of $7^-$ and $0^+$ phonons does not bring any modification as most of them lie above 15 MeV. Experimentally, a $0^+$ is observed in  $^{68}$Ni at $\sim 1.77$ MeV \cite{nndc}. However, the RQRPA calculation predicts the lowest $0^+$ state at $\sim 6$ MeV, and with a transition probability below the $5\%$ criterion mentioned above.
\\
The contribution of the unnatural parity modes was investigated systematically in the framework of quasiparticle-phonon model QPM \cite{PB.97}, which implies similar physical mechanisms of the formation of nuclear excited states as RQTBA. The role of such modes is known to be minor and, therefore, they are not included in our phonon space.
\\
Summarizing, our convergence study illustrates that although the Gamow-Teller strength, especially its low-energy part, is sensitive to the energy and angular momentum cut-off values, it nevertheless  shows clear saturation within the set of phonons with quantum numbers $J^\pi_{\mu} = 2^+$, $3^-$, $4^+$, $5^-$, $6^+$ and energies up to $\Omega_{max}=30$ MeV. The phonon model space truncated in this way was used in the numerical calculations discussed below.


\begin{figure*}
\centering
\resizebox{0.7\textwidth}{!}
{
  \includegraphics{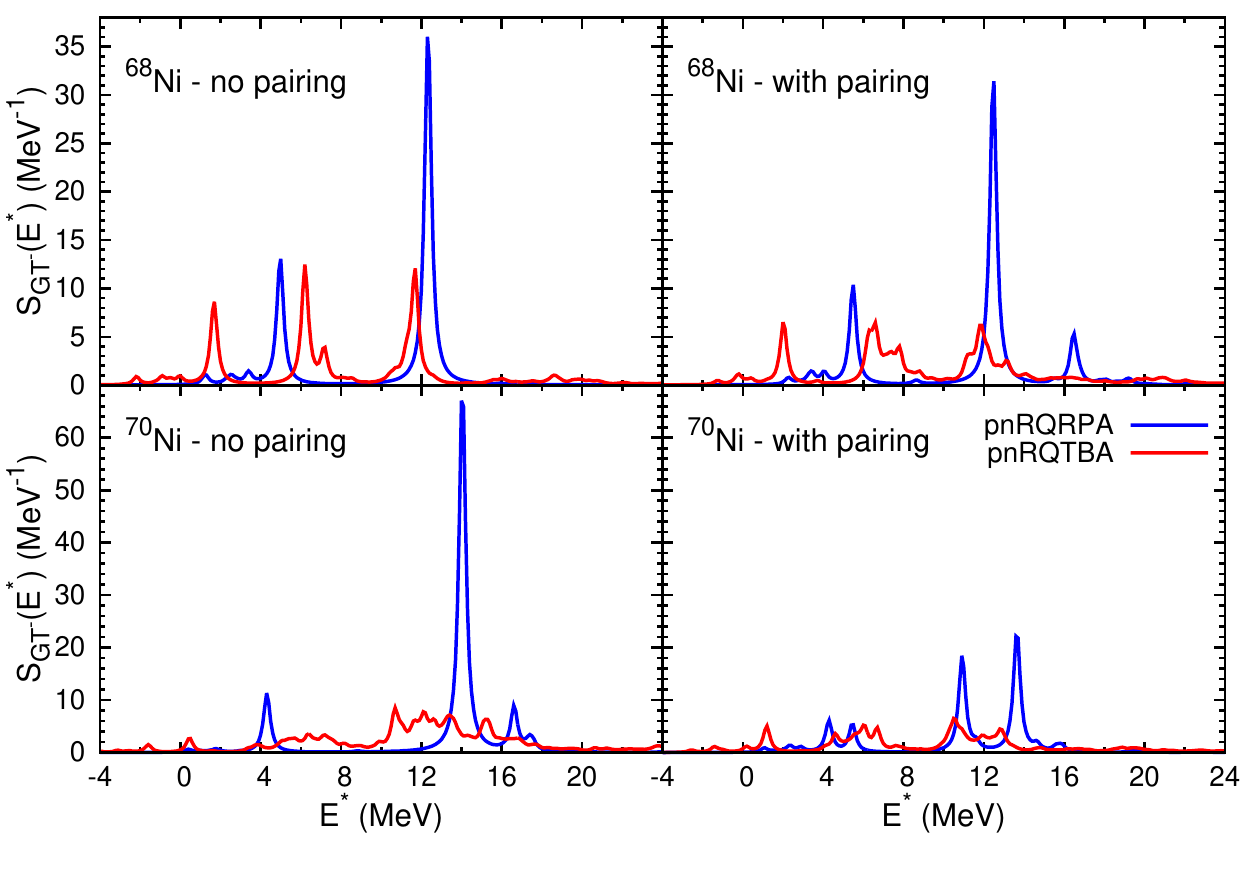}
}
\caption{(Color online) Gamow-Teller strength distributions $^{68}$Ni $\rightarrow$ $^{68}$Cu (top) and $^{70}$Ni $\rightarrow$ $^{70}$Cu (bottom), without (left) and with (right) superfluid pairing correlations, calculated with a smearing parameter $\Delta = 200$ keV. The excitation energy $E^*$ is expressed with respect to the ground-state energy of the parent nucleus $E_0$(Ni).}
\label{fig:strength_pair}      
\end{figure*}


\begin{figure*}
\centering
\resizebox{0.9\textwidth}{!}
{
  \includegraphics{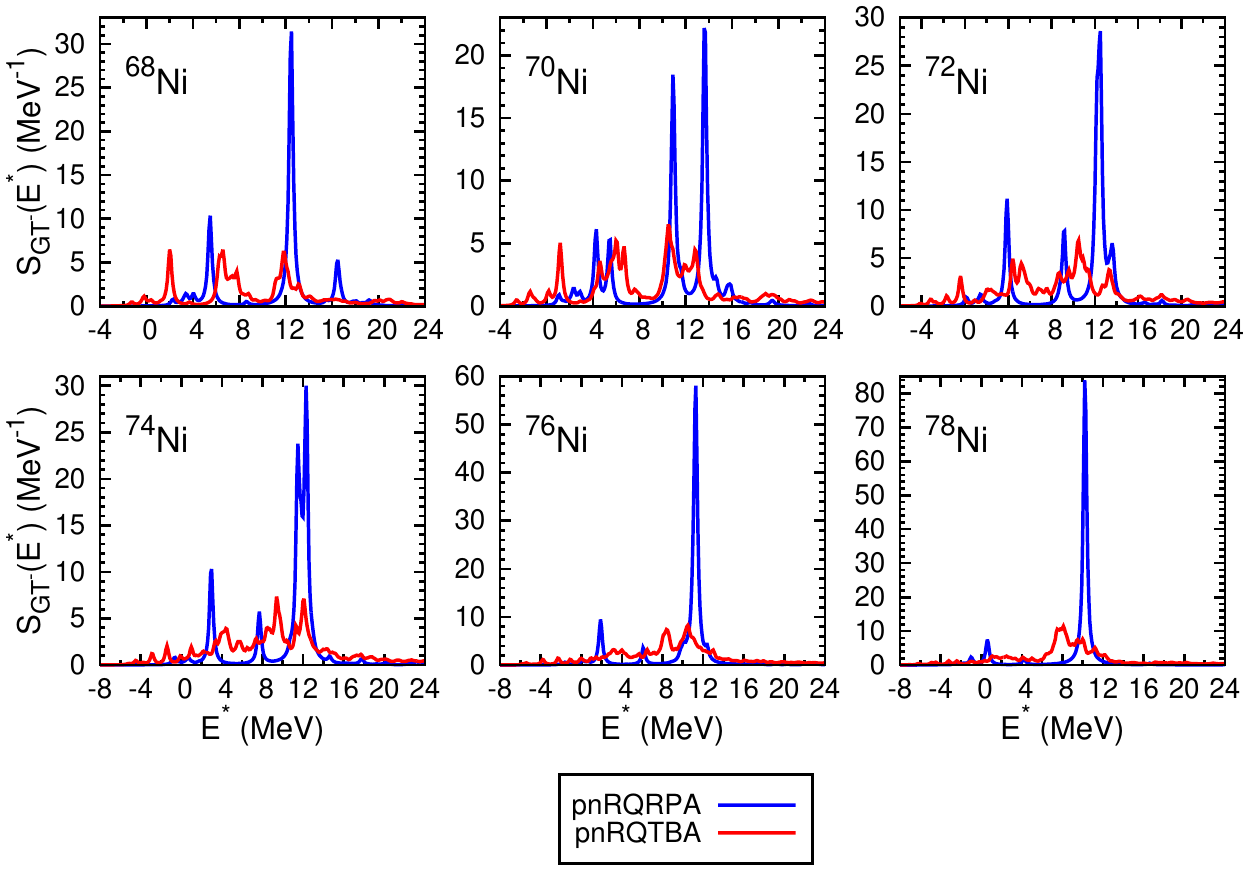}
}
\caption{(Color online) Gamow-Teller strength distributions Ni $\rightarrow$ Cu for the neutron-rich Nickel isotopic chain, calculated with a smearing parameter $\Delta = 200$ keV. The excitation energy $E^*$ is expressed with respect to the ground-state energy of the parent nucleus $E_0$(Ni).}
\label{fig:strength_all}      
\end{figure*}


\paragraph{Effect of superfluid pairing correlations}
In order to illustrate the effect of isospin-like superfluid pairing correlations, we first show in Fig. \ref{fig:strength_pair}, as an example, the strength distributions of the transitions $^{68}$Ni $\rightarrow$ $^{68}$Cu and $^{70}$Ni $\rightarrow$ $^{70}$Cu, and compare them to the distributions obtained without pairing correlations. The excitation energies $E^*$ are expressed with respect to the ground state of the parent Ni nucleus. In blue are the results obtained in the pn-RQRPA approximation, by neglecting the quasiparticle-vibration coupling amplitude $\boldsymbol{\Phi}(\omega)$ in the effective interaction (\ref{eq:eff_int}) of the BSE (\ref{eq:BSE_pn}), and in red are the strengths obtained when the coupling of quasiparticles to vibrations is introduced, i.e. when the full BSE (\ref{eq:BSE_pn}) is solved. Since the $^{68}$Ni nucleus has a closed neutron sub-shell at the mean-field level (the neutron 2p1/2 orbit is fully occupied), the inclusion of pairing correlations only has a small effect on the calculated strength. Within the pn-RQRPA approximation, the two main peaks at $\sim$ 5.0 and 12.3  MeV are very slightly shifted towards higher energies, by 500 and 200 keV respectively, and their intensity is a bit decreased, by 2.7 and 2.1 MeV$^{-1}$ respectively. The corresponding strength is transferred to a new peak that appears around 16.5 MeV. In $^{70}$Ni, the neutron 1g9/2 sub-shell is open at the mean-field level, with an occupation of 2 particles. As illustrated by Fig. \ref{fig:strength_pair}, the pairing correlations therefore have a much stronger effect on this nucleus. At the pn-RQRPA level, the main peak at $\sim 11.8$ MeV is split into two smaller ones around 10.9 and 13.6 MeV.  The fragmentation observed at the discrete pn-RQRPA level is attributed to Landau damping. Finally, the introduction of quasiparticle-phonon coupling brings an important fragmentation of the pn-RQRPA strength, which becomes distributed over a larger energy range, thus describing the major part of the spreading width of the resonances.

\paragraph{Gamow-Teller strengths in the Nickel isotopic chain} 
We now show in Fig. \ref{fig:strength_all} the strength distributions obtained for the full neutron rich even-even Nickel isotopic chain $^{68-78}$Ni.  Again we compare the results obtained with (pn-RQTBA) and without (pn-RQRPA) quasiparticle-vibration coupling. The fragmentation and spreading of the strength brought by this coupling is systematically observed along the Nickel chain, and the effect seems to be more important in the most neutron-rich isotopes $^{76-78}$Ni.

\subsubsection{Calculation of $\beta$-decay half-lives}
We now use the theoretical strength distributions to calculate the $\beta$-decay half-lives $T_{1/2}$ of the nuclei under consideration. They are calculated in the allowed GT approximation using the following formula \cite{Niu2015}:
\begin{equation}
\frac{1}{T_{1/2}} = \frac{g_a^2}{D} \int_{\Delta B}^{\Delta_{nH}} f(Z,\Delta_{np}-E) \, S(E) \, \mbox{d}E \; .
\label{eq:hlives}
\end{equation}
In Eq. (\ref{eq:hlives}), $S(E)$ is the strength distribution at excitation energy $E$ taken with respect to the ground state of the mother nucleus, and $f$ denotes the phase-space factor of the electron \cite{phase_space}, where $Z$ is the number of protons in the parent nucleus and $\Delta_{np}=1.293$ MeV is the mass difference between a neutron and a proton. The integration bounds $\Delta B$ and $\Delta_{nH}$ denote the binding energy difference $\Delta B = B(A,Z)-B(A,Z+1)$ between the mother and daughter nuclei, and the mass difference between a neutron and the Hydrogen atom  $\Delta_{nH}=0.78227$ MeV, respectively. Finally, the constant $D=6163.4$ s, and we consider the effective value for the weak axial coupling constant $g_a=1$.

\begin{figure}
\resizebox{0.5\textwidth}{!}
{
  \includegraphics{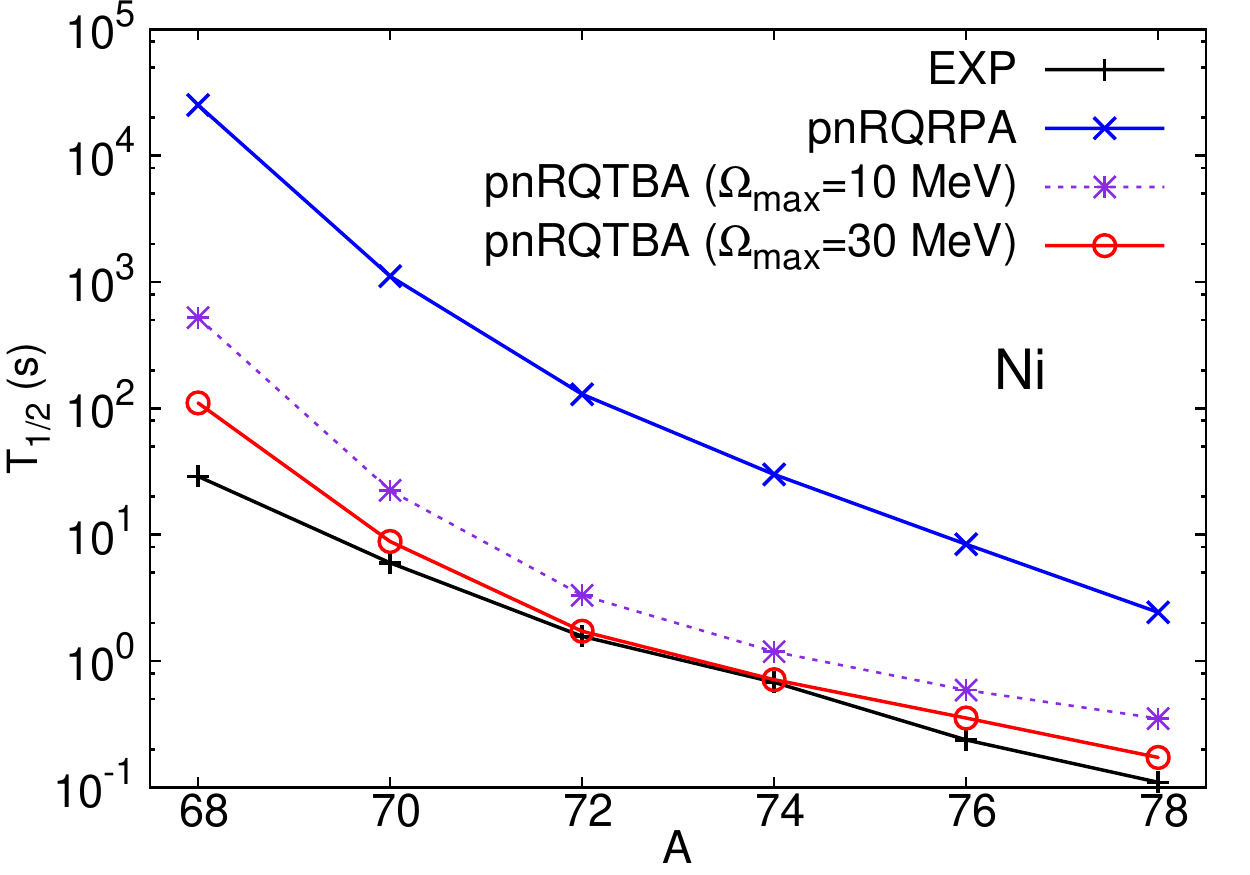}
}
\caption{(Color online) Theoretical half-lives compared to experiment, in seconds.}
\label{fig:hlives}      
\end{figure}

\begin{figure*}
\centering
\resizebox{0.8\textwidth}{!}
{
  \includegraphics{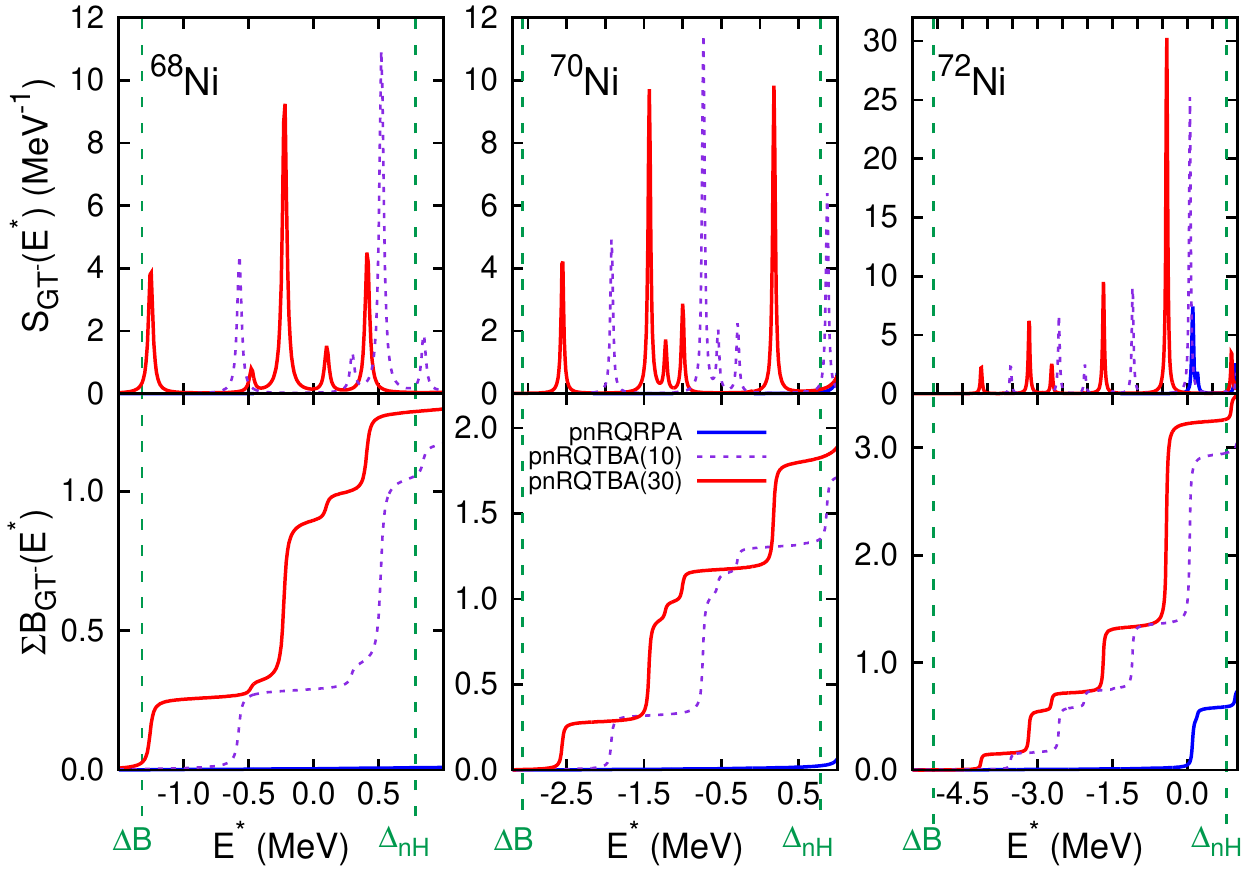}
}
\caption{(Color online) Top: Gamow-Teller ($GT^-$) strength distributions for $^{68-70-72}$Ni in the $Q_\beta$-window calculated with a smearing parameter $\Delta=20$ keV. Bottom: corresponding running integrated strength $\sum B_{GT^-}(E^*) = \int_{\Delta B}^{E^*} S(E) dE$. In the pn-RQTBA case, the phonon truncation energy $\Omega_{max}$ is indicated in parenthesis.}
\label{fig:strength_Qb_1}      
\end{figure*}

\begin{figure*}
\centering
\resizebox{0.8\textwidth}{!}
{
  \includegraphics{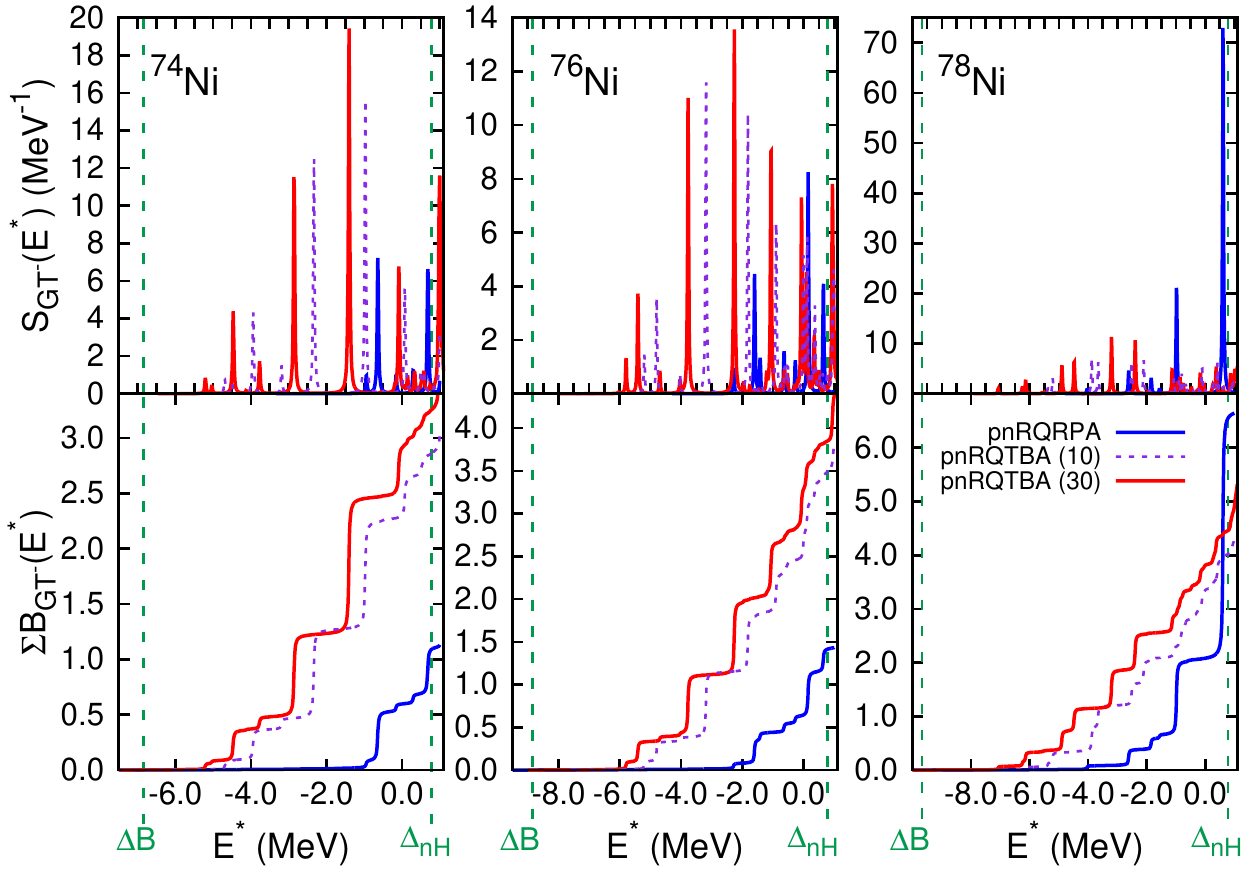}
}
\caption{(Color online) Same as Fig. \ref{fig:strength_Qb_1} for $^{74-76-78}$Ni.}
\label{fig:strength_Qb_2}      
\end{figure*}

\noindent In order to obtain a good precision on the theoretical half-lives, the strength functions in Eq. (\ref{eq:hlives}) are calculated with a smearing parameter $\Delta=20$ keV. 
The resulting $\beta$-decay half-lives are presented in Fig. \ref{fig:hlives}, and compared to the experimental ones \cite{nndc}. The half-lives obtained within the pn-RQRPA approximation, shown by blue crosses, overestimate the experimental data by up to three orders of magnitudes in the worst case ($^{68}$Ni). The introduction of the coupling between quasiparticles and phonons clearly improves the results. To illustrate the sensitivity of the half-lives to the phonon energy cut-off, we show the results obtained for two values of the phonon truncation energy $\Omega_{max}= 10$ MeV (purple crosses) and $\Omega_{max}= 30$ MeV (red circles). The difference between the results obtained in these two cases, is mainly due to the inclusion of the giant resonance region around $\sim 15-20$ MeV. In comparison to pn-RQRPA, the QPVC makes it possible to get one to two orders of magnitude closer to the experimental results, and a quite nice agreement with data is obtained when we include phonons up $\Omega_{max}=30$ MeV. We note that the lifetime obtained for $^{78}$Ni is in agreement with the results obtained in \cite{Niu2015}. That reference, however, showed a smaller value of $T_{1/2}$ for $^{68}$Ni, while treating this nucleus as doubly magic.
\\
For a better understanding of our results, we show in Figs. \ref{fig:strength_Qb_1} and \ref{fig:strength_Qb_2} the calculated strength in the $Q_\beta = \Delta_{nH} - \Delta B$ window, as well as the running integrated strength $\int_{-\Delta B}^{E^*} S(E) dE$, for $^{68}$Ni to $^{78}$Ni.
In the case of $^{68}$Ni and $^{70}$Ni, we note from Fig. \ref{fig:strength_Qb_1} the quasi-absence of strength in the $Q_\beta$ window at the pn-RQRPA level. In fact the integrated strength in this window is equal to $\sim 0.73 \times 10^{-2}$ and $2.51 \times 10^{-2}$ for these two nuclei, respectively. We emphasize that these finite values are actually artifacts due to the finite smearing parameter $\Delta$. No actual states are found in the $Q_\beta$-window for both of these isotopes and the small integrated strength is thus only produced by the tail of states located at higher energies. With a value $\Delta=0$ keV, $^{68}$Ni and $^{70}$Ni would therefore be predicted stable by the pn-RQRPA.
When QPVC is taken into account, finite strength appears at lower energy due to fragmentation effects, and a few peaks are now predicted within the $Q_\beta$-window. The values of the integrated strength, obtained with $\Omega_{max}= 30$ MeV, are equal to 
$\sim 1.26$ and $1.82$ for $^{68}$Ni and $^{70}$Ni, respectively, and the lifetimes calculated within pn-RQTBA for these two nuclei are thus predicted finite. 
For the three isotopes $^{72-74-76}$Ni, pn-RQRPA now shows some strength in the $Q_\beta$ window, while the distribution in pn-RQTBA is shifted to lower energies and the integrated strength is again found much larger.
The case of the doubly-magic nuclei $^{78}$Ni is quite different. Indeed, looking at Fig. \ref{fig:strength_Qb_2}, we note that pn-RQRPA actually predicts more strength in the $Q_\beta$ window than pn-RQTBA, mainly due to the peak located at $\sim 0.61$ MeV. The integrated GT strength at 0.78 MeV amounts to $6.55$ with pn-RQRPA and 
$4.46$ with pn-RQTBA ($\Omega_{max}=30$ MeV). However, pn-RQTBA predicts strength at lower energies than pn-RQRPA, where the phase-space factor in Eq. (\ref{eq:hlives}) is much larger. A smaller lifetime is then produced. Besides the total amount of strength, this illustrates the importance of the distribution of the strength within the integration window. \\ \\

Finally let us note that the study of the Nickel isotopic chain is in fact an interesting test case for the following reason. As mentioned in the introduction, in relativistic or non-relativistic calculations based on QRPA, an additional static proton-neutron pairing force is usually included and adjusted separately so to reproduce the lifetimes of each isotopic chain. However, it was concluded in Refs. \cite{Niksic} and \cite{Engel} that the T=0 pairing has very little effect in the Nickel chain because of the shell and sub-shell closures at $Z=28$ and $N=40$. The half-lives were therefore found not affected by the variation of the corresponding proton-neutron pairing strength.
This of course shows that there must be another mechanism responsible for the shift of the strength to lower energies. From the present study, it follows that the coupling between quasiparticle and collective vibrations provides the most important part of this effect.

\section{Conclusion}
\label{sec:conclu}
In the present work we have developed the formalism of the proton-neutron Relativistic Quasiparticle Time Blocking Approximation, 
which provides a description of charge-exchange excitations in closed and open-shell nuclei, while taking into account the coupling of quasiparticles to collective vibrations in a self-consistent way. This model is, on one hand, an extension of the pn-RQRPA \cite{Paar} by quasiparticle vibration coupling and, on the other hand, a generalization of the proton-neutron RTBA to superfluid pairing. A considerable strength of the method is that it is based on the fundamental meson-exchange nucleon-nucleon interaction and connects consistently the high-energy scale of heavy mesons, medium-energy range of pion and the low-energy domain of emergent collective vibrations. The latter two build up the effective interaction in the isospin-flip channel, in particular, the phonon-exchange part takes care of the retardation effects, which are of great importance for the spreading of the resonances, quenching and beta-decay rates with significant consequences for astrophysics and theory of weak processes in nuclei.     
\\
We have applied this approach to the calculations of Gamow-Teller strength distributions and beta-decay half-lives of neutron-rich Nickel isotopes. The results obtained  are very satisfactory. It is found that the quasiparticle-vibration coupling causes fragmentation of
the strength function of these nuclei, in both the giant resonance energy region and in the low-lying part of the distribution. Due to this fragmentation, a part of the strength was shifted to lower energies, including the $Q_\beta$ window. Consequently, this work illustrates the ability of the method to produce accurate $\beta$-decay half-lives, without introducing additional adjustable parameters, such as proton-neutron pairing strength, which is often used in the models on the QRPA level.
\\
Beta-decay is an important phase of r-process nucleosynthesis, alternating with the phase of radiative neutron capture (n,$\gamma$). As astrophysical modeling uses the rates of both $\beta$-decay and (n,$\gamma$) processes for producing elemental abundance distributions, it is very desirable to compute both of them within the same highly predictive framework. The present development allows for such calculations in combination with the existing RQTBA approach for nuclear response in the neutral channel. Similarly to the sensitivity of beta-decay rates to low-lying Gamow-Teller strength, the (n,$\gamma$) reaction rates are sensitive to fine details of the low-energy dipole strength \cite{LLL.08}. Until now RQTBA is only implemented numerically on the 2p2h level, while systematic calculations show that higher-order correlations are needed for high-precision description of the low-energy dynamics.  The recent advancement  of Ref. \cite{Litvinova2015} offers an extension of the response theory for both isospin-flip and non-isospin-flip excitations, showing a way of further resolving the fine spectral details, which could be the next conceptual step following the present work. Another opportunity to explore is the inclusion of the ground state correlations caused by QPVC mechanism \cite{Tselyaev} and isovector phonons formed by in-medium pion exchange as it has a sizable effect on nuclear shell structure \cite{L.16}. 
These and other aspects on nuclear spin-isospin response will be addressed by future efforts.

\section{Acknowledgements}
The authors thank Tomislav Marketin for helpful discussions and for providing a part of the code for pn-RQRPA matrix elements.
This work was supported by US-NSF Grants PHY-1404343 and PHY-1204486.

\end{document}